\documentclass[aps,prd,amsmath,amssymb,preprintnumbers,twocolumn]{revtex4}
\usepackage[a4paper,left=20mm,right=20mm,top=25mm,bottom=25mm]{geometry}
\usepackage{graphicx}
\usepackage{color}
\usepackage{subfigure}
\usepackage{pxfonts}
\usepackage{fancyhdr}
\usepackage{multirow}
\usepackage{tikz}

\definecolor{rossoCP3}{cmyk}{0,0.88,0.77,0.40}
\newcommand{\symb}{{\color{rossoCP3}{\vardiamondsuit}}}

\newcommand{\tr}[1]{\langle #1\rangle}
\newcommand{\cL}{\mathcal{L}}
\newcommand{\cO}{\mathcal{O}}

\begin{document}

\title{\Large \color{rossoCP3} SIMP Model at NNLO in Chiral Perturbation Theory}
\author{Martin Hansen$^\symb$}
\email{hansen@cp3.dias.sdu.dk}
\author{Kasper Lang\ae ble$^\symb$}
\email{langaeble@cp3.dias.sdu.dk} 
\author{Francesco Sannino$^\symb$}
\email{sannino@cp3.dias.sdu.dk} 
\affiliation{$^\symb${ CP}$^3${-Origins} and the Danish IAS, University of Southern Denmark, Campusvej 55, DK-5230 Odense M, Denmark.}

\begin{abstract}
We investigate the phenomenological viability of a recently proposed class of composite dark matter models where the relic density is determined by $3\to2$ number-changing processes in the dark sector. Here the pions of the strongly interacting field theory constitute the dark matter particles. By performing a consistent next-to-leading and next-to-next-to-leading order chiral perturbative investigation we demonstrate that the  leading order analysis cannot be used to draw conclusions about the viability of the model. We further show that higher order corrections substantially increase the  tension with phenomenological constraints challenging the viability of the simplest realisation of the strongly interacting massive particle (SIMP) paradigm.    \\[2mm]
{\footnotesize\it Preprint: CP$^3$-Origins-2015-025 DNRF90, DIAS-2015-25}
\end{abstract}

\maketitle

\section{Introduction}
Dark Matter accounts for  circa 85\%  of the matter in the universe, but besides from its cosmological abundance, very little is known about its nature. In a wide class of models, the relic abundance is generated via a thermal freeze-out in the early universe. Typically  $2\to2$ annihilation processes into e.g.~standard model particles keep dark matter in thermal equilibrium with the standard model bath until the annihilation processes drop  below the Hubble expansion rate. After this point in time the abundance of dark matter is essentially constant throughout the universe. This constitutes the weakly interacting massive particle (WIMP) paradigm. 

In a recent paper \cite{Hochberg:2014dra} the authors revived an alternative mechanism \cite{Carlson:1992fn,deLaix:1995vi} for achieving the observed dark matter relic density. Instead of using  $2\to2$ annihilation processes they assume that a dominant $3\to2$ number-changing process occurs in the dark sector involving strongly interacting massive particles (SIMPs). This process reduces the number of dark particles at the cost of heating up the sector. However, the presence of hot dark matter is problematic for structure formation, which means that at the time of freeze-out, the dark matter particles must to be in thermal equilibrium with the standard model ones. This, in turn, requires small couplings between the dark and the standard model sectors. In this way the energy from the dark sector can be transferred to the standard model via scattering processes.

The coupling between the two sectors allows for direct and indirect detection, while the large self-interactions can play a role in structure formation, by solving the {\it core vs.~cusp} problem \cite{deBlok:2009sp}. Compared to the WIMP paradigm, where the dark matter particles typically are believed to be around the TeV scale, this model can yield dark matter particles with masses around a few 100 MeVs. This is an interesting alternative to the WIMP paradigm given the fact that current experiments are putting  substantial constraints on this old paradigm. These constraints can, however, be alleviated or even be completely offset within the recently proposed {\it safe dark matter} paradigm \cite{Sannino:2014lxa}. 
  
A follow-up paper \cite{Hochberg:2014kqa} introduced a realization of the SIMP mechanism based on an underlying strongly coupled sector described via chiral perturbation theory. In this set-up the pions constitute the dark matter particles and a key role is played by the time-honoured Wess-Zumino-Witten (WZW) term \cite{Wess:1971yu,Witten:1983tw,Witten:1983tx}. The WZW term is non-vanishing in theories where the coset space of the symmetry breaking pattern has a non-trivial fifth homotopy group. This topological term introduces a 5-point pion interaction, making it an ideal candidate for the $3\to2$ annihilation process. In QCD, for example, the term describes the annihilation of two kaons into three pions. In this paper we shall only be concerned with the symmetry breaking pattern $SU(2N_f) \to Sp(2N_f)$ for $N_f = 2$. The simplest realization of this breaking pattern comes from an underlying $Sp(2)=SU(2)$ gauge group, but in general it can be realized for any $Sp(N_c)$ gauge group. The actual pattern of chiral symmetry breaking depends on the number of flavours, colors and matter representation. A comprehensive study of the conformal window for $Sp(N_c)$ gauge groups can be found in \cite{Sannino:2009aw}.   Lattice simulations have further demonstrated that such an underlying dynamics truly leads to the expected pattern of chiral symmetry  \cite{Lewis:2011zb} with the spectrum of the composite spin-one resonances first computed in \cite{Lewis:2011zb,Hietanen:2014xca,Hietanen:2013fya}. 
 
The computations performed in \cite{Hochberg:2014kqa} make use of the first non-vanishing order in the chiral expansion for the $3 \to 2$ and $2\to 2$ processes. For the $3\to 2$ annihilation process it is one order higher than the related $2\to2$ self-interaction process. We will demonstrate that it is important to analyse the physical results via a consistent next-to-leading (NLO) and next-to-next-to-leading order (NNLO) chiral perturbative treatment. We will, in fact, show that the leading order analysis is phenomenologically unreliable because it is outside the range of convergence. The important higher order corrections substantially increase the range of convergence of the theory, and therefore the phenomenological reach. Within our controllable analysis we discover that higher order corrections increase the tension with respect to the phenomenological constraints making it hard for the SIMPlest realisation to be phenomenologically viable. 

\section{Consistent Setup}
Chiral perturbation theory is an expansion in powers of the pion mass and momentum. Throughout this paper we will use the standard terminology LO, NLO and NNLO when referring to $\cO(p^2)$, $\cO(p^4)$ and $\cO(p^6)$, respectively. Furthermore, beyond leading order it is well known that a consistent chiral perturbation analysis requires the introduction of a number of low-energy constants   that uniquely identify the underlying strongly coupled theory. The NLO low-energy constants, relevant for the $2\to2$ process, are denoted here by N-LECs. At NNLO both processes contain low-energy constants which we denote by NN-LECs.

Since the $3\to2$ process emerges naturally at the four-derivative level, it does not have a leading order contribution. On the other hand the $2\to2$ process does have leading order contributions, introducing a potential mismatch in power counting between the two processes. In our set-up the solution to the Boltzmann equation (which depends on the thermally averaged cross section for the $3\to2$ process) returns a value for the pion decay constant, as a function of the pion mass, such that the expected relic density is obtained. These values are subsequently substituted into the cross section for the $2\to2$ self-interaction, which is constrained by observations of e.g.~the bullet cluster \cite{Markevitch:2003at}. Technically, the mismatch arises because the two processes, unless computed to the same order, disagree on the definition of the physical pion mass $m_\pi$ and decay constant $f_\pi$ which at NLO can be written as:
\begin{align}
 m_\pi^2 &= m^2\left[1 + \frac{m^2}{f^2}(a_mL+b_m) + \cO\left(\frac{m^4}{f^4}\right)\right] \\
 f_\pi &= f\left[1 + \frac{m^2}{f^2}(a_fL+b_f) + \cO\left(\frac{m^4}{f^4}\right)\right]
\end{align}
Here $m$ and $f$ are the bare parameters from the leading order Lagrangian, $a_i$ and $b_i$ are combinations of N-LECs and $L$ is a log term defined in equation \eqref{eq:log}. The amplitude should, however, be expressed in terms of the physical quantities by rewriting the bare quantities in terms of the physical ones. From this argumentation it is clear that a consistent calculation requires both processes to be calculated at the same order.

In practice it means that one has to go beyond leading order when computing the $2\to 2$ process, and this is what we will do here. The additional benefit of extending the computation of the $2\to2$ processes to NLO  is that we will also be able to estimate, and extend, the convergence of the perturbative expansion towards the physically relevant  regime. To check the stability of the perturbative expansion we also estimate the NNLO corrections.  Because we are interested only in a preliminary estimate of the NNLO corrections we retain the direct contributions from the loop diagrams and neglect the  finite contributions stemming from the NN-LECs. We will show that the NNLO results are already much closer to the NLO, than the NLO are to the LO for a much wider range of pion masses. 

\section{Scattering and Relic Density}
The Boltzmann equation  for the $3\to2$ process reads 
\begin{equation}
 \label{boltzmann}
 \dot{n} + 3Hn = -(n^3-n^2n_{eq})\langle\sigma v^2\rangle_{3\to2},
\end{equation}
where $n(t)$ is the pion number density and $H$ is the Hubble constant. As argued in \cite{Hochberg:2014dra} contributions from the $2\to2$ annihilation into standard model particles can be neglected, because they are sub-dominant. To solve the differential equation we rewrite it in terms of the dimensionless quantities $Y=n/s$ and $x=m_\pi/T$ with $s$ being the entropy and $T$ the temperature. For the entropy and matter density we used the definitions
\begin{equation}
 \rho = \frac{\pi^2}{30}g_e(T)T^4,\qquad s = \frac{2\pi^2}{45}h_e(T)T^3,
\end{equation}
where $g_e$ and $h_e$ are the effective degrees of freedom contributing to the entropy and density, respectively. In this notation, the differential equation becomes
\begin{equation}
 \frac{dY}{dx} = -\sqrt{\frac{4\pi^5g_*}{91125G}}\frac{m_\pi^4}{x^5}(Y^3-Y^2Y_{eq})\langle\sigma v^2\rangle_{3\to2},
\end{equation}
where $g_*(x)$ combines information from both $g_e(T)$ and $h_e(T)$ into a single function for the effective degrees of freedom (the function is different from the function used in the $2\to2$ case). The equilibrium function $Y_{eq}$ can be written as
\begin{equation}
 Y_{eq} = \frac{45N_\pi x^2}{4\pi^4 g_e(m_\pi/x)}K_2(x),
\end{equation}
where $K_2(x)$ is the modified Bessel function of the second kind and $N_\pi$ is the number of pions. The thermally averaged scattering cross section used in \cite{Hochberg:2014kqa} reads
\begin{equation}
 \langle\sigma v^2\rangle_{3\to2} = \frac{75\sqrt{5}}{512\pi^5x^2}\frac{m_\pi^5}{f_\pi^{10}}\frac{N_c^2}{N_\pi^3},
\end{equation}
where $N_c$ is the number of colors and our definition of the pion decay constant $f_\pi$ differs by a factor of one-half compared to the original paper. The value of $f_\pi$ is chosen in such a way that the solution to the Boltzmann equation gives the expected relic density. An estimate of the higher order corrections to the thermally-averaged cross section can be found in the appendix.
\begin{figure*}[t]
\begin{center}
 \includegraphics[width=0.49\linewidth]{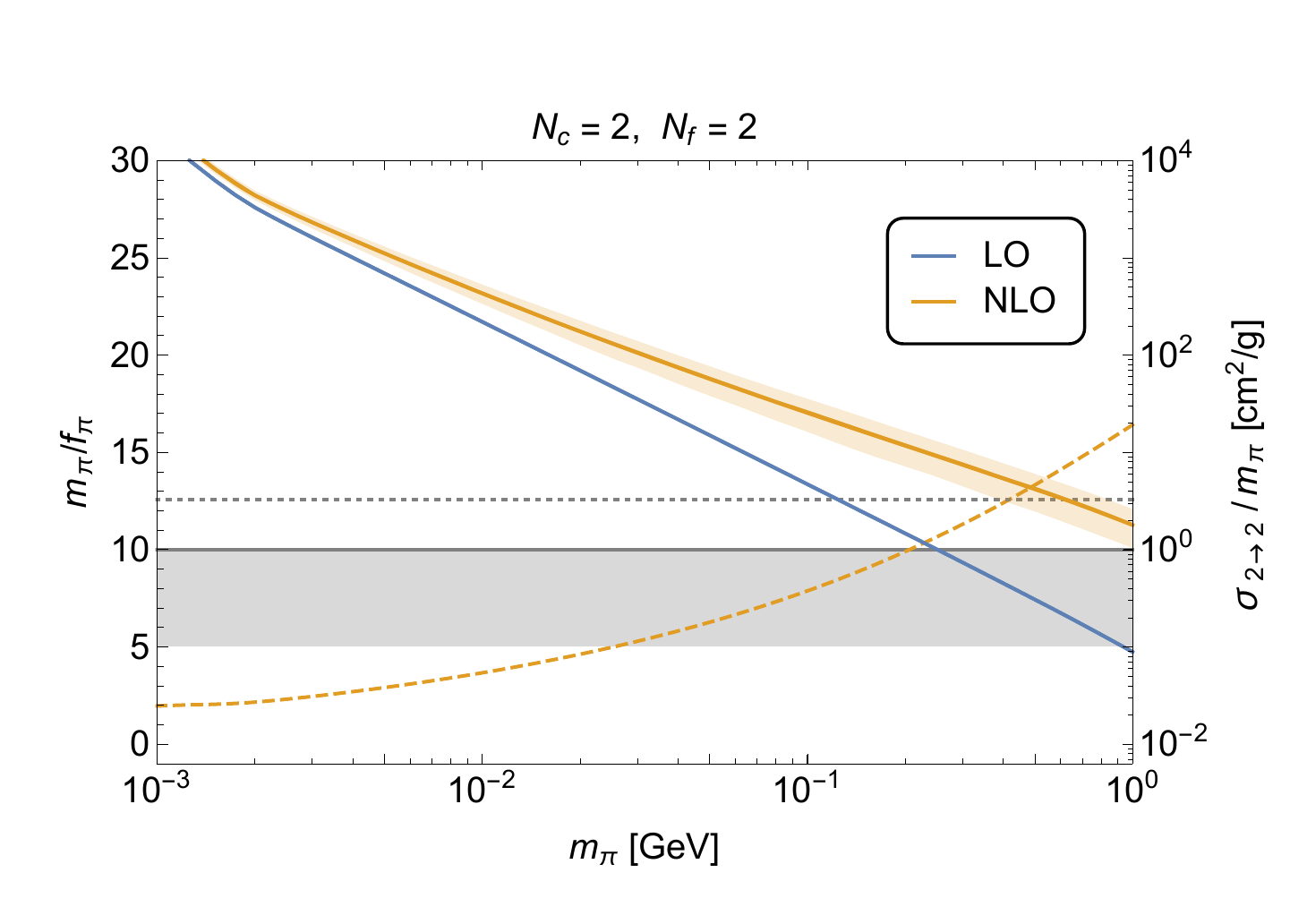}
 \includegraphics[width=0.49\linewidth]{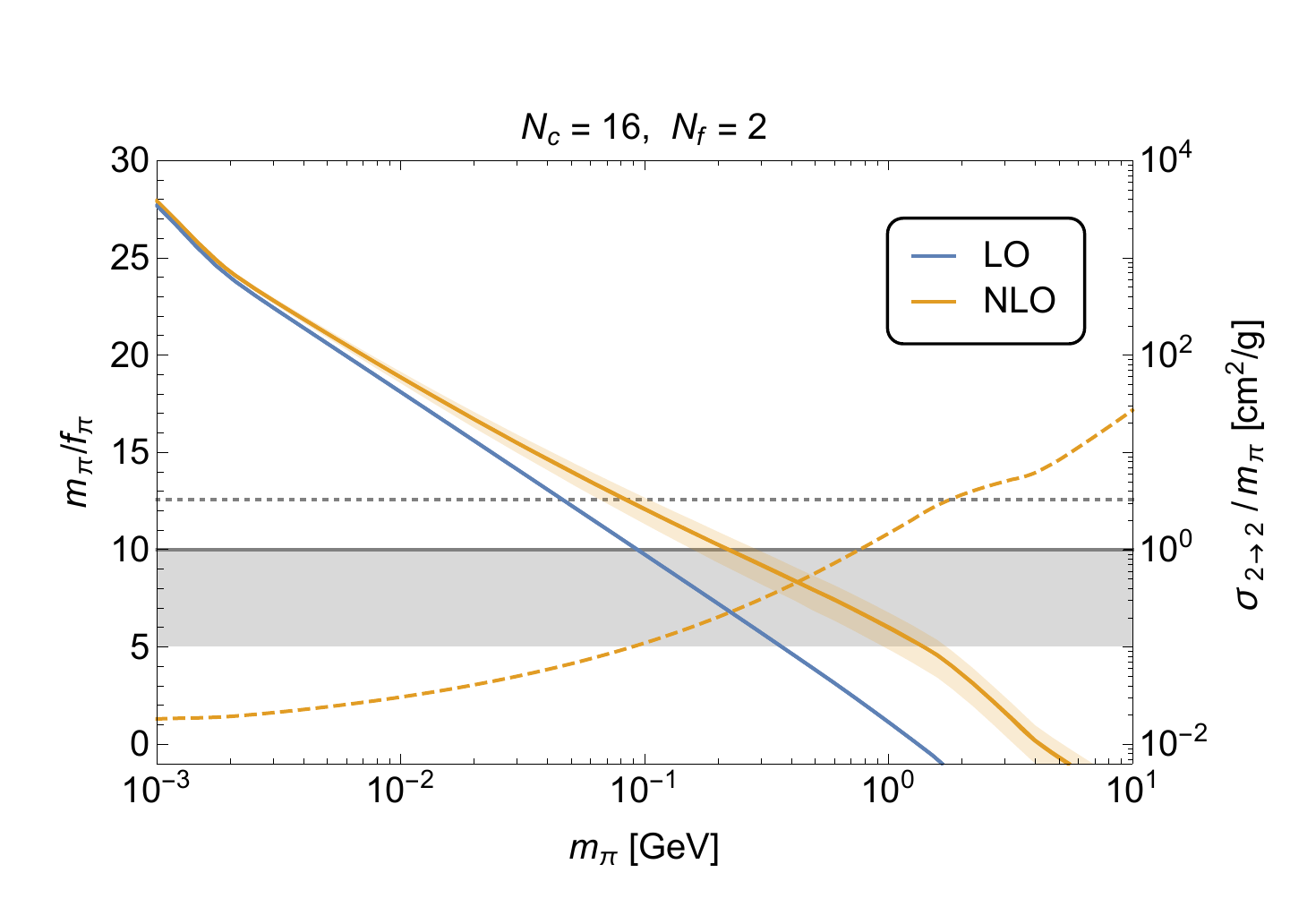}
 \caption{\textit{Dashed} lines belong to the left axis and \textit{solid} lines to the right axis. The dashed orange line is the solution $m_\pi/f_\pi$ to the Boltzmann equation and the dashed horizontal line is the upper perturbative limit $m_\pi/f_\pi=4\pi$. The two solid lines are the cross section for the $2\to2$ self-interactions at LO (blue) and NLO (orange). The band on the orange line is the uncertainty from the low-energy constants (N-LECs). The solid grey band is the upper limit on the self-interactions.}
 \label{fig:NLO}
\end{center}
\end{figure*}

\subsection{Pion-pion Scattering}
To determine the pion-pion scattering amplitude we make use of the non-linearly realized  chiral Lagrangian. We adopt the notation of \cite{Bijnens:2011fm} together with the results from the NNLO. In fact, we have independently performed the NLO computations for the breaking pattern $SU(2N_f)\to Sp(2N_f)$ and shown that the earlier published results \cite{Bijnens:2011fm} must be amended. To keep the work self-contained we briefly outline the computational setup.

Let $G$ be the global flavor symmetry of the vector-like fermions transforming according to a given representation of the underlying gauge dynamics, and let $H$ be the stability group after spontaneous symmetry breaking. The associated Goldstone boson manifold $G/H$ is parametrized by 
\begin{equation}
 u = \exp\left(\frac{i}{\sqrt{2}f}X^a\phi^a\right),
\end{equation}
where $X^a$ are the broken generators, normalized as $\tr{X^aX^b}=\delta^{ab}$ where $\tr{\cdot}$ denotes trace in flavor space. The quantity $u$ transforms as 
\begin{equation}
 u \to g u h^\dagger = h u g^\dagger,
 \label{constraint}
\end{equation}
with $g\in G$ and $h\in H$. Here $g$ is a space-time independent element of $G$ while $h$ is a space-time dependent element of $H$ that must satisfy the constraint equation \eqref{constraint}. The chiral Lagrangian  quantities that transform homogeneously under the stability group $H$ are
\begin{align}
 u_\mu &= i(u^\dagger(\partial_\mu-ir_\mu)u - u(\partial_\mu-il_\mu)u^\dagger), \\
 \chi_\pm &= u^\dagger\chi u^\dagger \pm u\chi^\dagger u.
\end{align}
In the first expression $r_\mu$ and $l_\mu$ are external currents that can be set to zero in the computation of the pion-pion scattering amplitude. In the second expression $\chi$ is a spurion field that formally ensures chiral symmetry invariance at every step in the computation. Only at the very end it is replaced by its expectation value $\chi = m^2\mathbb{I}$ which explicitly breaks chiral symmetry. The pion mass is then written as $m^2=2B_0m_q$ where $B_0$ is related to the underlying chiral condensate and $m_q$ is the new quark mass. In this notation the LO Lagrangian  is
\begin{equation}
 \cL_2 = \frac{F^2}{4}\tr{u_\mu u^\mu + \chi_+},
\end{equation}
and the relevant part of the NLO Lagrangian is
\begin{align}
\begin{split}
 \cL_4
 &= L_0\tr{u_\mu u_\nu u^\mu u^\nu}
 + L_1\tr{u_\mu u^\mu}\tr{u_\nu u^\nu} \\
 &\quad+ L_2\tr{u_\mu u_\nu}\tr{u^\mu u^\nu}
 + L_3\tr{u_\mu u^\mu u_\nu u^\nu} \\
 &\quad + L_4\tr{u^\mu u_\mu}\tr{\chi_+}
 + L_5\tr{u^\mu u_\mu\chi_+} \\
 &\quad+ L_6\tr{\chi_+}^2 +  L_7\tr{\chi_-}^2
 + \tfrac{1}{2}L_8\tr{\chi_+^2+\chi_-^2}.
\end{split}
\end{align}
Because the NLO Lagrangian represents the most general Lagrangian it is possible to absorb the one-loop divergences by an appropriate renormalization of the low energy constants.  We use the modified $\overline{\mathrm{MS}}$ scheme where
\begin{equation}
 L_i = L_i^r - \frac{\Gamma_i}{32\pi^2}R,
\end{equation}
with
\begin{equation}
 R = \frac{2}{\epsilon} + \log(4\pi) - \gamma_E + 1.
\end{equation}
\begin{figure*}[t]
\begin{center}
 \includegraphics[width=0.49\linewidth]{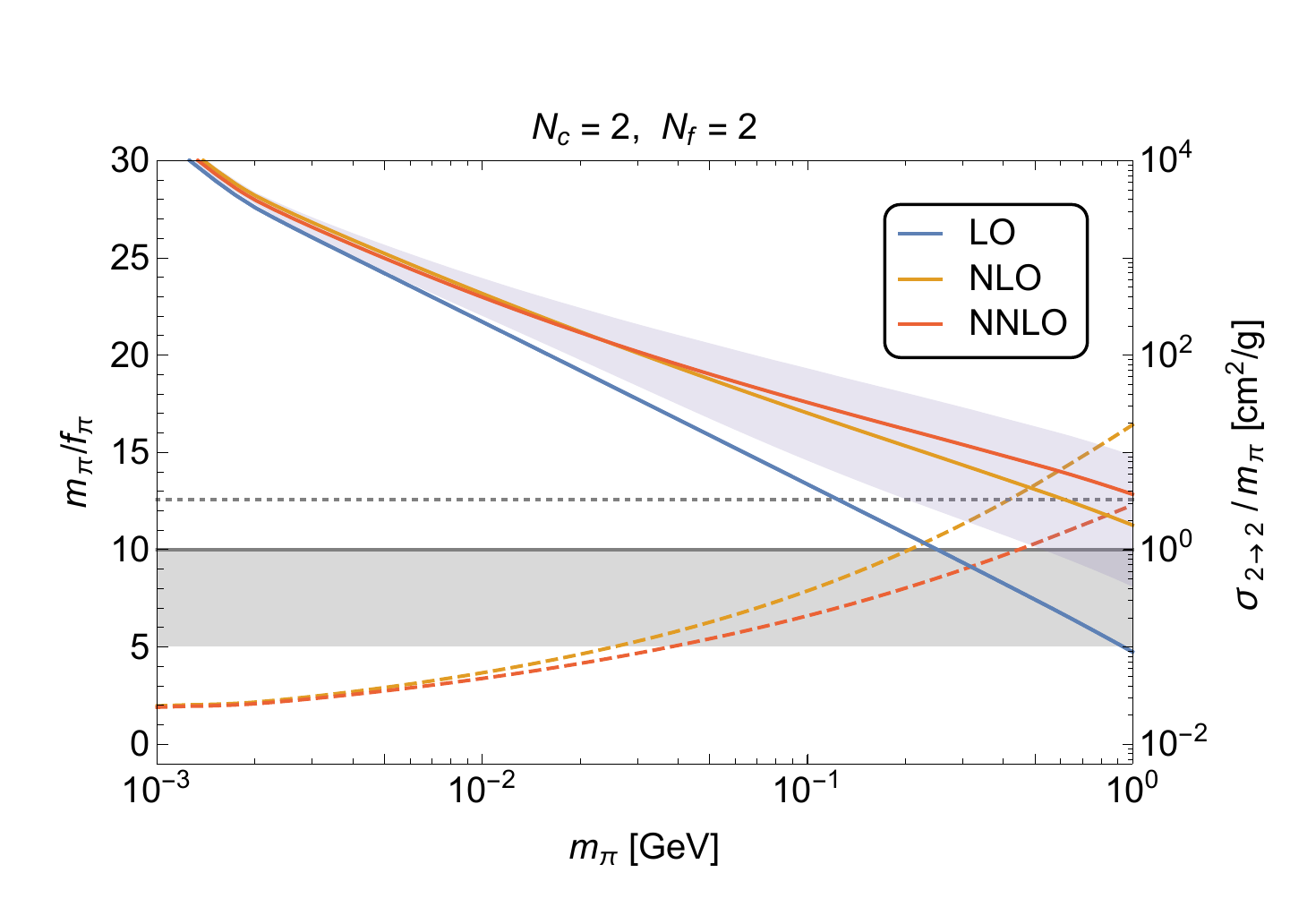}
 \includegraphics[width=0.49\linewidth]{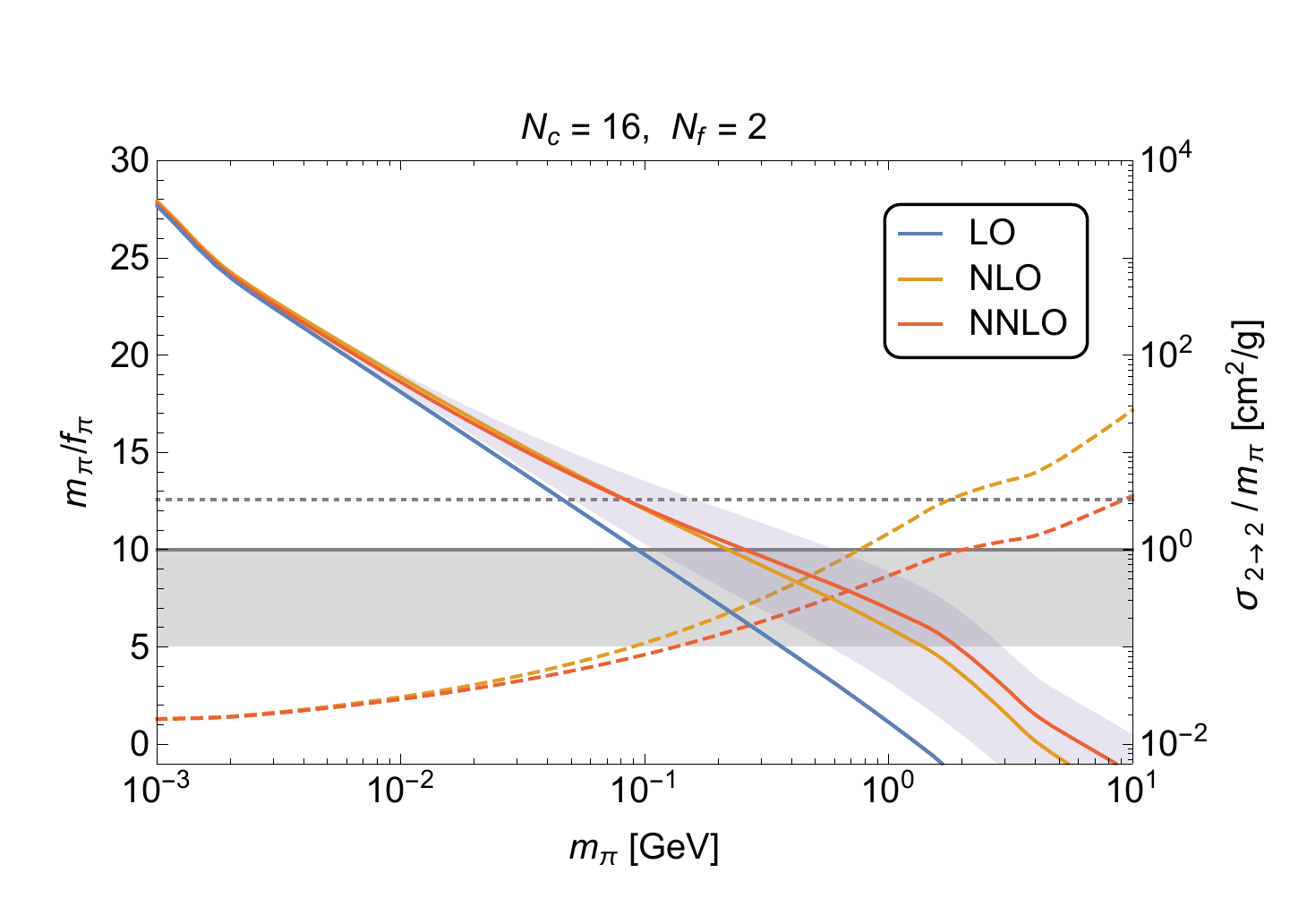}
 \caption{\textit{Dashed} lines belong to the left axis and \textit{solid} lines to the right axis. The red dashed line is the NNLO solution $m_\pi/f_\pi$ to the Boltzmann equation, the orange dashed is the NLO, and the dashed (grey) horizontal line is the upper perturbative limit $m_\pi/f_\pi=4\pi$. The three solid lines are the cross section for the $2\to2$ self-interactions at LO (blue), NLO (orange) and NNLO (red). The purple band is the uncertainty from the low-energy constants (N-LECs). The solid grey band is the upper limit on the self-interactions.}
 \label{fig:NNLO}
\end{center}
\end{figure*}
Here $\epsilon=4-d$ and $\gamma_E=-\Gamma'(1)$ is the Euler-Mascheroni constant. It should be noted that the renormalized coefficients $L_i^r$ will depend on the energy scale $\mu$ introduced by dimensional regularization. To simplify the results later on we use the following short-hand notation for the log terms arising from the loop diagrams.
\begin{equation}
 L = \pi_{16}\log\left(\frac{m_\pi^2}{\mu^2}\right),\qquad \pi_{16} = \frac{1}{16\pi^2}.
 \label{eq:log}
\end{equation}
At NLO there are three new diagrams contributing to the amplitude for pion-pion scattering (see \cite{Bijnens:2011fm} for details and diagrams) and at NNLO there are twelve additional diagrams. The scattering amplitude can be written as
\begin{align}
\begin{split}
 T(s,t,u) &= \xi^{abcd}B(s,t,u) + \xi^{acbd}B(t,u,s) \\
 &+ \xi^{adbc}B(u,s,t) + \delta^{ab}\delta^{cd}C(s,t,u) \\
 &+ \delta^{ac}\delta^{bd}C(t,u,s) + \delta^{ad}\delta^{bc}C(u,s,t),
\end{split}
\end{align}
where $\xi^{abcd}$ is a group theoretical factor that depends on the breaking pattern and the number of flavors. For the case $SU(4) \to Sp(4)$ all relevant quantities can be found in the appendix.

\subsection{Topological term}
The topological term can be written in a compact local form as a five-dimensional integral \cite{Wess:1971yu,Witten:1983tw,Witten:1983tx} known as the Wess-Zumino-Witten (WZW) term 
\begin{equation}
 S_{WZW} = \frac{N_c}{240\pi^2}\int_0^1 d\alpha\int d^4x~\epsilon^{abcde}\tr{u_au_bu_cu_du_e},
\end{equation}
were $\alpha$ is the fifth spacetime coordinate. We furthermore redefine the two quantities
\begin{align}
 u &= \exp\left(\frac{i\alpha}{\sqrt{2}f}X^a\phi^a\right), \\
 u_a &= i(u^\dagger\partial_a u - u\partial_a u^\dagger),
\end{align}
with $\partial_a$ being a five-dimensional derivative. Ordinary Minkowski space is now defined as the surface of the five-dimensional space where $\alpha=1$. Furthermore the pre-factor of the WZW term contains direct information about the underlying gauge dynamics. The gauged version of this term and its generalization to include left and right vector sources can be found in \cite{Duan:2000dy}. 

At NLO there is only one diagram for the $3\to2$ process while at the NNLO there are three more diagrams. However, only two of these diagrams are non-zero in the approximation of vanishing NN-LECs.
\begin{figure*}[t]
\begin{center}
 \includegraphics[width=0.49\linewidth]{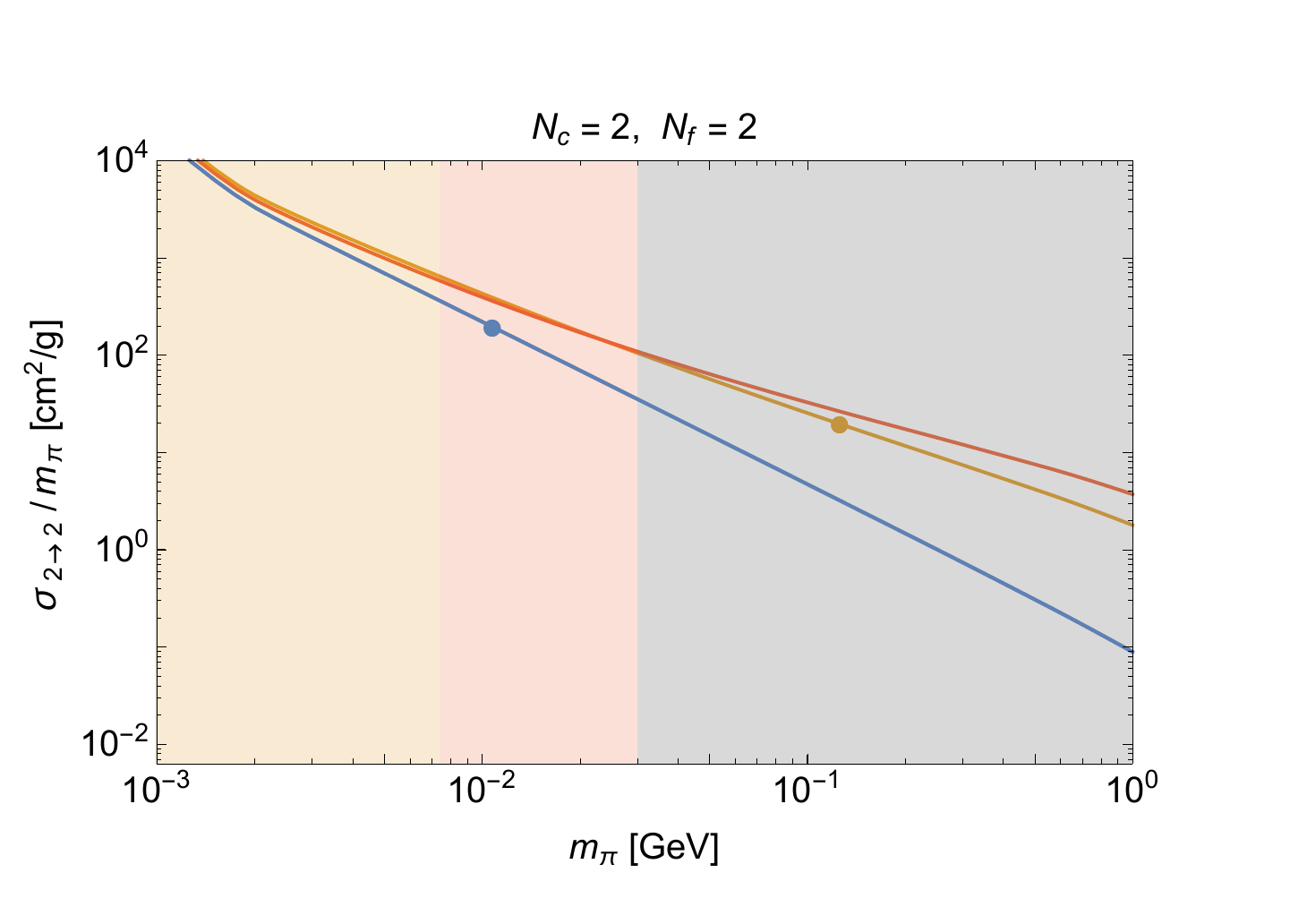}
 \includegraphics[width=0.49\linewidth]{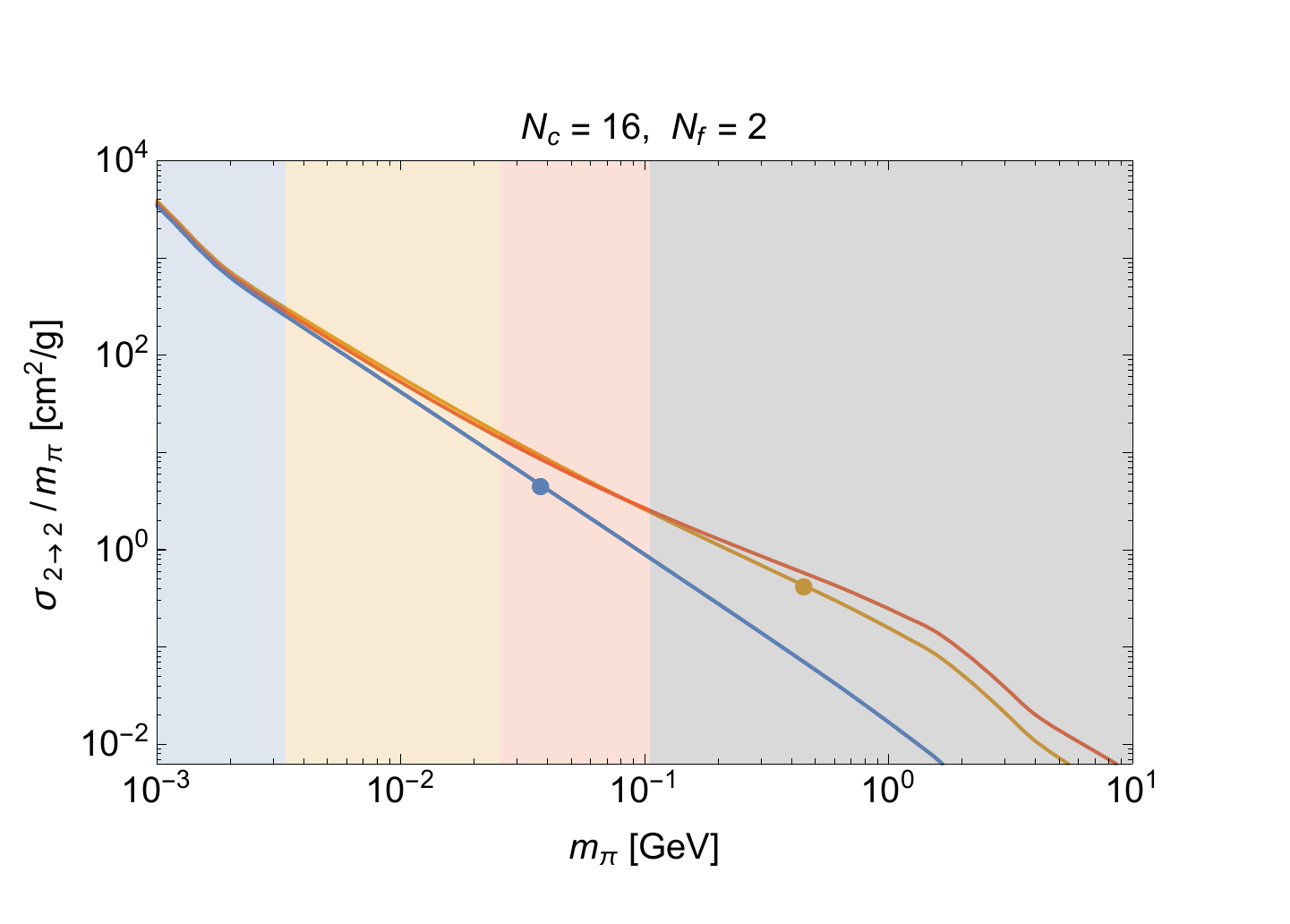}
 \caption{The shaded regions indicate where the different orders in perturbation theory can be trusted up to around 20\%. The blue line is LO, orange line is NLO and red line is NNLO. The dots show for, each order, when the next order corrections are of the same magnitude. The estimates are obtained neglecting the LECs.  }
 \label{fig:FLAG}
\end{center}
\end{figure*}

\section{Results}
We are now ready to discuss the physical results both at the NLO and NNLO. 

At the NLO we retain the N-LECs that uniquely specify the underlying strongly coupled theory. These constants are  expected to be a number of order unity suppressed by a factor $\pi_{16}$, defined in equation \eqref{eq:log}. To determine the LECs one can either use experiments, when available (as for QCD), or perform lattice simulations when the underlying theory is known \cite{Lewis:2011zb,Hietanen:2013fya,Hietanen:2014xca}. For QCD the N-LECs are known \cite{Bijnens:2014lea} and most of them are of the order $\cO(10^{-4})$. Interestingly this is an order of magnitude smaller than na\"ively expected. Therefore to estimate the N-LECs we randomize their values using a Gaussian distribution with spread  $5\times10^{-4}$ and zero mean. By averaging over a large number of samples we arrive at a NLO band that reflects the size of the contributions from the N-LECs.

In figure~\ref{fig:NLO} we plot the solution to the Boltzmann equation (dashed orange line) and the cross section for the $2\to2$ self-interactions (solid lines) at LO (blue line) and NLO (orange line). The band on the orange line is the estimated effect from the low-energy constants. The horizontal dashed line is the upper perturbative limit $m_\pi/f_\pi=4\pi$. The cross section for the self-interactions is constrained from above by the solid grey line. The contraint from the bullet cluster \cite{Markevitch:2003at} is $\sigma/m_\pi \lesssim 1$ cm$^2$/g but simulations of halo structures \cite{Zavala:2012us,Rocha:2012jg} suggest that the limit might be closer to $\sigma/m_\pi \lesssim 0.1$ cm$^2$/g. This uncertainty is reflected in the grey band.

The model is phenomenologically relevant when the dashed orange line is below the perturbative limit and the self-interactions are below the solid grey line. 

However, for $N_c=2$ these two criteria are never met simultaneously when the NLO corrections are properly taken into account. The results demonstrate that the NLO corrections are crucial to establish the phenomenological viability of the model. The right panel of figure~\ref{fig:NLO} shows that by substantially increasing the number of colors one may still hope to meet the phenomenological criteria although in this case a more careful use of the large-$N_c$ expansion is needed.

Because we assume small couplings to the standard model, there is also a lower limit on the dark matter mass of around 10 MeV. In fact, lighter masses will change the effective number of neutrino species \cite{Boehm:2013jpa} which is in tension with the Planck data \cite{Ade:2013zuv}.

Since, for the region of phenomenological interest, the NLO corrections are quite sizeable (especially for small values of $N_c$) we also estimate the NNLO corrections. To consistently include these higher order corrections, we re-determined the solution to the Boltzmann equation using the thermally averaged cross section
\begin{equation}
 \langle\sigma v^2\rangle_{3\to2}^{NNLO} = \langle\sigma v^2\rangle_{3\to2}^{NLO}\left(1 + \frac{m_\pi^2}{f_\pi^2}(a_wL+b_w)\right),
\end{equation}
at NNLO. The term $a_wL$ comes from the loop corrections whereas the last term $b_w$ would contain the NN-LECs if we had included them. As it turns out, the solution is relatively insensitive to the value of $b_w$ assuming it is of order $\cO(\pi_{16})$. This means that setting the low-energy constants to zero in the  amplitude for the $3\to2$ process is a reasonable approximation. For the self-interactions we estimate the effects from just the N-LECs in the same way as before.

In figure~\ref{fig:NNLO} we show the solution to the Boltzmann equation (dashed red line) and the $2\to2$ scattering cross section (solid red line) at NNLO. There are two important points to make, namely that (a) the solution $m_\pi/f_\pi$ decreases, making the investigation more amenable to a perturbative analysis, while (b) the cross section seems to increase further, strengthening the tension with the self-interaction constraints. If the solution did not decrease one would have seen a more dramatic increase in the NNLO cross section. This underlines the importance of a consistent calculation. Despite the small difference between the NLO and NNLO cross sections, we observe a more prominent effect from the LECs as indicated by the purple band. There are two reasons for the increased uncertainty. It is partly due to the fact that the N-LECs enter more frequently in the NNLO result, but also because the squared norm of the NLO amplitude enters in the NNLO cross section. Because this quantity is strictly positive, the average of a Gaussian distribution deviates from zero.

In figure~\ref{fig:FLAG} we provide regions of validity where the different orders in perturbation theory can be trusted up to around 20\%. The dots show where each order breaks down because the corrections from the next order are of equal magnitude. For the NNLO case the values are rough estimates from extrapolating via the previous orders. In the grey region the theoretical uncertainty of the NNLO expansion is more than 20\%. We therefore estimate that for $N_c=2$ perturbation theory at NNLO is reasonable up to around 30 MeV. However, besides the aforementioned constraint coming from the effective number of neutrino species for masses below 10 MeV, this region is excluded by the constraint on the self-interactions. Even by allowing the NNLO treatment to be used beyond this point, the model cannot easily meet the phenomenological constraints. It is clear from the figure that the claimed phenomenologically relevant region \cite{Hochberg:2014kqa} is far beyond LO perturbation theory.

For the case of $N_c=16$ one has control up to around 110 MeV at NNLO, but here the theory is still in tension with the constraints on self-interactions. However, after this point the theory meets the constraints for plausible combinations of low-energy constants.

One can introduce additional breaking of the flavour symmetry which, in principle, can reduce the $2\to2$ scattering amplitude \cite{Hochberg:2014kqa}, but it will not change the range of convergence for the different perturbative orders.

\section{Conclusion}
By consistently using chiral perturbation theory we studied the phenomenological viability of an interesting class of composite dark matter models. In these models the relic density of the dark pions is achieved via $3\to2$ number-changing processes making use of the WZW term. We determined both the $3\to2$ and the $2\to2$ processes to the NLO and NNLO order in the chiral expansion and showed that higher order corrections substantially affect the LO result of \cite{Hochberg:2014kqa} for the $2\to 2$ processes. At the NLO and NNLO we have shown that  the SIMPlest models \cite{Hochberg:2014kqa}, with a moderate number of underlying colors, are at odds with phenomenological constraints. However, for sufficiently many colors, or via additional breaking of the flavor symmetry, it is still possible to find small regions in the parameter space consistent with phenomenological constraints.

Our results are also of immediate interest for different realizations of composite dark matter \cite{Ryttov:2008xe} and/or composite (Goldstone) Higgs \cite{Appelquist:1999dq,Cacciapaglia:2014uja,Arbey:2015exa} featuring the same pattern of chiral symmetry breaking.  Furthermore first principle lattice simulations of the underlying dynamics \cite{Lewis:2011zb,Hietanen:2013fya,Hietanen:2014xca} are able to provide crucial information on the low-energy constants \cite{Arthur:2014zda}.  

\acknowledgments
The CP$^3$-Origins centre is partially funded by the Danish National Research Foundation, grant number DNRF90. MH is funded by a Lundbeck Foundation Fellowship.

\appendix
\section{Scattering Amplitude}
The cross section used for the plots has been evaluated at the scattering threshold where $s=4m_\pi^2$ and $t=u=0$. In this limit the cross section reads
\begin{equation}
 \sigma_{2\to2} = \frac{|T|^2}{128\pi N_\pi^2m_\pi^2} \ ,
\end{equation}
with $T$ the amplitude of the process. The group theoretical factor in the scattering amplitude for the $SU(4)\to Sp(4)$ pattern of chiral symmetry breaking is
\begin{equation}
 \xi^{abcd} = \frac{1}{2}(\delta^{ab}\delta^{cd} - \delta^{ac}\delta^{bd} + \delta^{ad}\delta^{bc}),
\end{equation}
with $\{a,b,c,d\}=1,\dots,5$. The functions $B(s,t,u)$ and $C(s,t,u)$ evaluated on the kinematical threshold read: \\[2mm]
\textit{LO} -- All terms should be multiplied by $m_\pi^2/f_\pi^2$.
\begin{align*}
 B(s,t,u) &= 1 \\
 B(u,s,t) &= -1 \\
 C(s,t,u) &= 0 \\
 C(u,s,t) &= 0
\end{align*}
\textit{NLO} -- All terms should be multiplied by $m_\pi^4/f_\pi^4$.
\begin{align*}
 B(s,t,u) &= \tfrac{1}{2}(2\alpha_1 + 32\alpha_4 + 7\pi_{16}) \\
 B(u,s,t) &= (\alpha_1 + 4\alpha_2 + 16\alpha_3 - \pi_{16}) \\
 C(s,t,u) &= \tfrac{1}{4}(4\beta_1 + 16\beta_2 + 64\beta_3 + 17\pi_{16}) \\
 C(u,s,t) &= (\beta_1 + 16\beta_4 + \pi_{16})
\end{align*}
\textit{NNLO} -- All terms should be multiplied by $m_\pi^6/f_\pi^6$.
\begin{align*}
 B(s,t,u) &= \omega_1\pi_{16}^2 + \omega_2\pi_{16} + 16\gamma_4 + \gamma_1 \\
 B(u,s,t) &= \omega_3\pi_{16}^2 + \omega_4\pi_{16} + 64\gamma_5 + 16\gamma_3 + 4\gamma_2 + \gamma_1 \\
 C(s,t,u) &= \omega_5\pi_{16}^2 + \omega_6\pi_{16} + 64\delta_5 + 16\delta_3 + 4\delta_2 + \delta_1 \\
 C(u,s,t) &= \omega_7\pi_{16}^2 + \omega_8\pi_{16} + 16\delta_4 + \delta_1
\end{align*}
Here we made use of crossing symmetry which dictates that $B(t,u,s)=B(s,t,u)$ and $C(t,u,s)=C(u,s,t)$. The definitions of $\omega_i$ are given by
\begin{align*}
 \omega_1 &= \tfrac{35}{36}\pi^2 - \tfrac{55}{6}\\
 \omega_2 &= 16(L_0^r+L_3^r+3L_5^r+L_8^r) \\
          &\quad+ 128(L_1^r+L_2^r-L_4^r+L_6^r)-\tfrac{23}{2}L \\
 \omega_3 &= \tfrac{7}{9}\pi^2 - \tfrac{10}{3} \\
 \omega_4 &= 32(L_0^r+L_3^r-L_5^r+L_8^r) \\
          &\quad-128(L_1^r+L_2^r-L_4^r+L_6^r)+\tfrac{13}{2}L \\
 \omega_5 &= \tfrac{20}{3} - \tfrac{14}{9}\pi^2 \\
 \omega_6 &= 152(L_0^r+L_3^r+L_8^r)-8L_5^r\\
          &\quad+416(L_1^r+L_2^r+L_4^r+L_6^r)-\tfrac{393}{8}L \\
 \omega_7 &= \tfrac{1}{18}\pi^2 - \tfrac{5}{3} \\
 \omega_8 &= -32(L_0^r+L_3^r-L_5^r+L_8^r)-2L
\end{align*}
The remaining constants $\alpha_i$, $\beta_i$, $\gamma_i$ and $\delta_i$ can be found in \cite{Bijnens:2011fm}. For the breaking pattern $SU(2N_f)\to Sp(2N_f)$ considered here, we amended the expression for $\alpha_2$ given in \cite{Bijnens:2011fm} where the sign in front of the $\pi_{16}$ term should be negative. We furthermore set all LECs at NNLO to zero.  We choose the renormalization scale $\mu^2=4m_\pi^2$  to agree with the energy scale of the process. 

\section{Thermal Average}
The thermally averaged cross section for the $3\to2$ process at NLO is 
\begin{equation}
 \langle\sigma v^2\rangle_{3\to2}^{NLO} = \frac{75\sqrt{5}}{512\pi^5x^2}\frac{m_\pi^5}{f_\pi^{10}}\frac{N_c^2}{N_\pi^3},
\end{equation}
and at NNLO reads
\begin{equation}
 \langle\sigma v^2\rangle_{3\to2}^{NNLO} = \langle\sigma v^2\rangle_{3\to2}^{NLO}\left(1 + \frac{m_\pi^2}{f_\pi^2}(a_wL+b_w)\right),
\end{equation}
with the following definitions
\begin{align}
 a_w &= -\frac{77}{6} \simeq -12.83 \\
 b_w &= \frac{5\sqrt{5}}{288\pi^2}\log\left(\frac{9+\sqrt{45}}{9-\sqrt{45}}\right) - \frac{7}{96\pi^2} \simeq 1.83\cdot10^{-4}
\end{align}
If we had included the NN-LECs they would appear in $b_w$.

\end{document}